# Inelastic collisions of ultracold triplet Rb$_2$ molecules in the rovibrational ground state

Björn Drews[1], Markus Deiß[1], Krzysztof Jachymski[2], Zbigniew Idziaszek[2] and Johannes Hecker Denschlag[1]

[1]Institut für Quantenmaterie and Center for Integrated Quantum Science and Technology IQ$^{ST}$, Universität Ulm, D-89069 Ulm, Germany, [2]Faculty of Physics, University of Warsaw, Pasteura 5, 02-093 Warsaw, Poland

Exploring inelastic and reactive collisions on the quantum level is a main goal of the developing field of ultracold chemistry. We present first experimental studies of inelastic collisions of metastable ultracold triplet molecules in the vibrational ground state. The measurements are performed with nonpolar Rb$_2$ dimers which are prepared in precisely-defined quantum states and trapped in an array of quasi-1D potential tubes. In particular, we investigate collisions of molecules in the absolute lowest triplet energy level where any inelastic process requires a change of the electronic state. Nevertheless, we find similar decay rates as for collisions between rotationally or vibrationally excited triplet molecules and they are close to the rates for universal reactions. As anticipated theoretically, the measured decay rate constants vary considerably when confinement and collision energy are changed. This might be exploited to control the collisional properties of molecules.

Recent advances in the preparation of ultracold molecular samples in well-defined quantum states[1–7] sparked increasing interest in studying molecular collisions and chemical reactions on a pure and fundamental level[8–12]. Such experiments were first carried out with highly excited molecules[13–24] (also in the context of Efimov physics[25–30]), which can vibrationally relax in a collision. For molecules in the vibrational ground state this decay is barred, but other interesting reaction paths remain. First investigations of reactive or inelastic loss of singlet molecules in the rovibronic ground state with polar KRb and RbCs have recently been carried out[5,31–34]. Besides the electronic ground state $X^1\Sigma_g^+$, the metastable triplet state $a^3\Sigma_u^+$ is of special interest for collision experiments. If collisionally long-lived, such triplet molecules would allow for many interesting applications, such as tunable Feshbach resonances, due to their sizeable magnetic moment. To investigate chemical reactions between cold molecules, optical lattices are a convenient testbed as they allow for either isolating the molecules from each other or letting them collide. In addition optical lattices offer the possibility to control the dimensionality of the scattering process and to tune the interaction[35–38]. Based on this approach, it has been shown that strong inelastic collisions induce correlations and can inhibit particle loss in a molecular sample, a manifestation of the quantum Zeno effect[15,34].

In this work we present the first measurements on ultracold collisions of metastable triplet molecules that are internally in their lowest energy state. Specifically, we use $^{87}$Rb$_2$ dimers of the $a^3\Sigma_u^+$ state which are in the lowest hyperfine level of the rovibrational ground state. As reference measurements, we carry out collision experiments with rotationally excited molecules (rotational quantum number $R=2$) and with vibrationally highly excited Feshbach molecules. Initially, the dimers are prepared in a cubic 3D optical lattice with at most a single molecule per lattice site (Methods). By quickly ramping down one of its directions, the lattice is converted into an array of quasi-1D potential tubes (see Fig. 1a). Subsequently, molecules within the same tube collide with tunable relative energies on the order of $\mu$K$\times k_B$, far above the Tonks gas regime of the work of Syassen et al.[15]. A single tube is typically filled with only a few molecules and can be considered as a closed few-body system since tunnelling between the tubes is negligible.

Whenever a collision between two molecules is inelastic or reactive, enough energy is released to expel all products out of the lattice. This is because the lattice depths are comparatively shallow, being on the order of about 10 $\mu$K$\times k_B$, which corresponds to about 210 kHz $\times h$. After a given interaction time $t$ we measure the total number of remaining Rb$_2$ molecules $N(t)$ and the width of the whole cloud $\sigma_z^c(t)$ along the tubes. From the observed decay of $N$ we conclude that a large part of the molecules is already lost in the first possible collision, quite independently of the internal vibrational excitation. Using a simple model we extract the decay rate constants and investigate how they depend on the confinement of the potential tubes. These results are then compared to predictions of a quantum defect model[39,40].

In the following, we first discuss the collision experiments with Rb$_2$ Feshbach molecules as this will help us analysing the data for the $v=0$ states. Figure 2a shows three data sets of $N(t)$ corresponding to different confinements of the potential tubes. We observe a strong loss of molecules within the first tens of milliseconds. It is striking that the decay takes place in a step-wise fashion which, after about 100 ms, gives way to a much slower exponential decay with a corresponding time constant of more than one second. This slow exponential decay is similar to the one we observe in a deep 3D optical lattice (Fig. 2a, inset) which is due to background gas collisions and spontaneous photon scattering[41]. Figure 2b shows that the width $\sigma_z^c$ oscillates synchronously to the steps.

We interpret these dynamics as follows: as one direction of the 3D lattice is quickly ramped down (within 400 $\mu$s), the particles are suddenly released from their individual lattice sites into 1D tubes. Along these tubes there is a harmonic confinement with trap frequency $\omega_z$ due to the Gaussian intensity profile of the 2D lattice laser beams. The molecules will synchronously undergo an oscillatory motion along this direction with a period $T=2\pi/\omega_z$, while being strongly confined transversally. As a result, the width of the observed cloud $\sigma_z^c$ oscillates with $2\omega_z$ (see Fig. 2b). Whenever the cloud is small and dense, the probability for molecular encounters and losses is increased. On the other hand, if the cloud is large the dimers are separated from each other and $N(t)$ stays almost constant. Thus, the longitudinal oscillatory motion explains the step-like decay of the molecules. Once all inelastic collisions have taken place the fast losses stop and the remaining signal corresponds to single molecules in the tubes. This regime is typically reached after $50-100$ ms.

In order to model the molecular decay in a quantitative way, we first reconstruct the distribution of the molecules in our 3D optical lattice. The molecules are initially produced from a Gaussian-shaped cloud of ultra-cold atoms in the optical lattice via magnetic Feshbach ramping (Methods). We assume Poisson statistics for the atomic occupation of each individual lattice site. Only lattice sites that are occupied by exactly two atoms will finally be occupied by a single molecule. All other lattice sites will end up empty. Thus, a given total atom number and cloud size fixes the initial molecular distribution, see Fig. 1b. We have verified that the predicted cloud size and total number of the Feshbach molecules indeed agree with our measurements (Methods). Taking the molecule distribution of Fig. 1b the histogram of the resulting filling of the 1D tubes is depicted in Fig. 1c. The average occupation is about 2.2. The histogram helps us to gain additional insights. If we assume inelastic two-body collisions to be the only source of particle loss, then the number of molecules remaining after a decay time of $\approx 100$ ms equals the number of tubes with an odd initial filling. Evenly occupied ones, by contrast, end up empty. By comparing the experimentally measured fraction of remaining particles to this prediction we can check the model for consistency (Supplementary Information, Sec. 3).

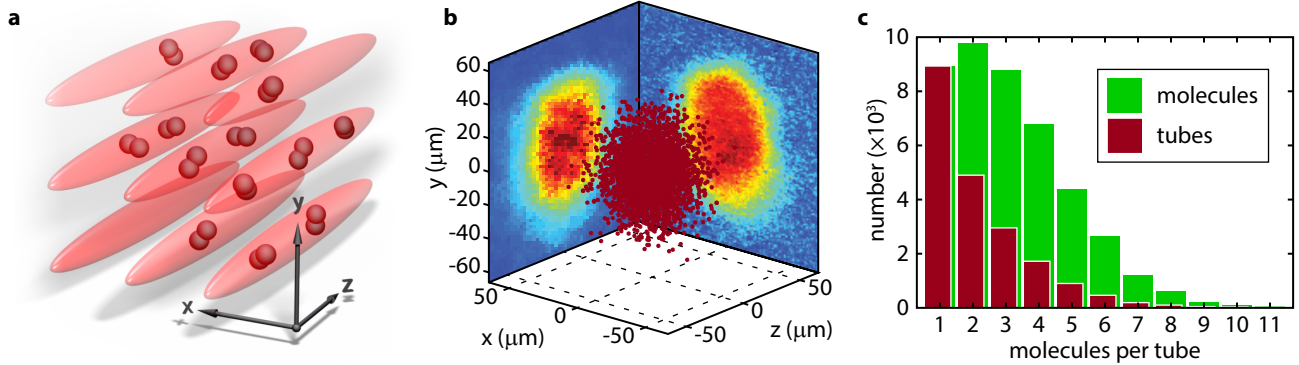

**Figure 1** | **Initial spatial distribution of molecules. a**, Illustration of molecules confined in an array of quasi-1D traps (potential tubes) within which they can collide. **b**, Random-generated Gaussian-shaped molecular cloud (red dots) of $4.3 \times 10^4$ Feshbach molecules which matches the observed molecular distribution as determined by the absorption images in the $y$, $z$- and the $x$, $y$-plane. **c**, Histogram of the molecular occupation of tubes as inferred from **b**. The red bars count the number of tubes with a given filling, while the green bars count the total number of molecules located in these tubes.

We now discuss the dynamics within a single 1D tube in more detail. Starting with a classical treatment, we consider each molecule as a point-like particle which is initially localised and at rest in a single lattice site of the 3D lattice. The molecules are all released exactly at the same time and consequently meet in the centres of the tubes precisely at $t = T/4 = \pi/(2\omega_z)$. At that particular moment the molecular cloud size ideally vanishes along the $z$-direction, inelastic collisions take place and the total molecule number decreases abruptly. In our measurements, however, we do not observe such abrupt steps, but rather smoothed ones (see Fig. 2a). We explain this fact by a non-vanishing initial velocity distribution of the molecules as a result of the Heisenberg uncertainty relation. Therefore, we leave the classical picture of point-like particles and rather describe each molecule $i$ as a 1D quantum mechanical wave packet $|\Psi_i(z,t)|^2 \propto \exp\left[-(z-\chi_i(t))^2/(\sqrt{2}\sigma_z^{wp}(t))^2\right]$, which is centred at $\chi_i(t)$ and has the width $\sigma_z^{wp}(t)$ (cf. Fig. 3a). This leads to a finite width in the molecule's velocity distribution, initially given by $\Delta v(0) \approx \hbar/(m\sigma_z^{wp}(0))$, where $m$ is the molecular mass. As $t$ progresses, $\chi_i(t)$, $\sigma_z^{wp}(t)$ and $\Delta v(t)$ oscillate with $\omega_z$, $2\omega_z$ and $2\omega_z$, respectively (Supplementary Information, Sec. 5). The dynamics for three particles in a tube is depicted in Fig. 3b (upper part). As expected, the collision times are now somewhat smeared out, however, every particle will still pass by every other one in the 1D tube within $T/2$ (assuming that no collisions oc-cur in the meantime). Whenever two molecules collide, there is a certain likelihood for an inelastic process. A collision between two molecules $i$ and $j$ is possible as long as the spatial overlap $\eta_{ij} = \int |\Psi_i|^2 |\Psi_j|^2 dz$ of their wave packets (cf. Fig. 3b, lower part) does not vanish. In order to describe the loss rate in a single 1D tube we choose the following ansatz (see Supplementary Information, Sec. 1, for a derivation)

$$\left\langle \frac{dN}{dt} \right\rangle = -\overline{\mathcal{K}}_{1D} \sum_{i \neq j} \eta_{ij}(t) \mathcal{F}(t), \quad (1)$$

where we sum twice over all molecular pairs $(i, j)$. $\mathcal{F}(t) = 4/\int |\Psi_i(z,t)\Psi_j(z',t) + \Psi_i(z',t)\Psi_j(z,t)|^2 dz\, dz'$ is a normalization factor. $\overline{\mathcal{K}}_{1D}$ is the mean decay coefficient. We note that in the limit of high tube occupation numbers $\overline{\mathcal{K}}_{1D}$ describes the dynamics of a 1D thermal gas with density $n$ via the rate equation $\dot{n} = -2\overline{\mathcal{K}}_{1D}\, n^2$.

We use Eq. (1) to analyse the measured decay curves and to extract $\overline{\mathcal{K}}_{1D}$. For this, we fit the solution of Eq. (1) (Methods) to the measurements. Indeed, the only free fitting parameter in our model is $\overline{\mathcal{K}}_{1D}$. This is because we can experimentally determine the initial spatial distribution of the molecules, and the initial width $\sigma_z^{wp}(t=0)$ of the molecular wave packets is derived by calculating the release dynamics from the 3D optical lattice into the 1D tubes (Supplementary Information, Sec. 5). For a given molecular sample we assume $\sigma_z^{wp}(t=0)$ to be identical for all particles. The values of $\sigma_z^{wp}(t=0)$ for the present work range between 0.17 to 0.26 $a_{lat}$, where $a_{lat} = 532$ nm is the lattice constant. Generally, the width $\sigma_z^{wp}(t=0)$ determines how smoothed out the decay steps are, while $\overline{\mathcal{K}}_{1D}$ determines their relative heights. Therefore, a variation in $\sigma_z^{wp}(t=0)$ does not have a strong influence on the extracted rate coefficient $\overline{\mathcal{K}}_{1D}$. As can be seen in Fig. 2a the fitted curves agree well with the measurements. Our results for the decay rates are $\overline{\mathcal{K}}_{1D} = (2.8, 3.6, 5.5)$ mm/s for the corresponding radial trap frequencies of $\omega_r = 2\pi \times (8.2, 11.6, 17.0)$ kHz, respectively (see also Fig. 5).

We now get back to Fig. 2b. A strong damping of the oscillations of the cloud size $\sigma_z^c(t)$ along the tubes is observed, which at first might be unexpected for 1D systems where thermalization is generally suppressed[42]. We attribute the damping mainly to the fact that the reaction rate increases with collision energy (as will be shown below) and therefore particles with higher kinetic energy are lost faster. A further discussion also of other possible contributions to the damping is provided in the Supplementary Information (Sec. 4). In addition, we would like to note that the observed oscillating cloud size in Fig. 2b is generally larger than expected from our model calculations. This is due to limitations in the *effective* imaging resolution (Methods).

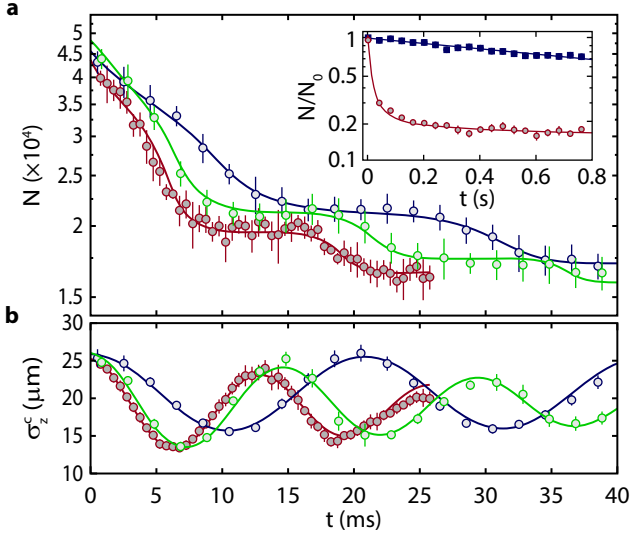

**Figure 2 | Dynamics of Feshbach molecules for various confinements.**
**a**, Decay of the molecule number $N(t)$ in an ensemble of quasi-1D traps with longitudinal trap frequencies $\omega_z = 2\pi \times (23.6, 33.4, 49.0)$ Hz, and transverse frequencies $\omega_r = 2\pi \times (8.0, 11.6, 17.2)$ kHz (blue, green, red). The continuous lines are fitted model calculations. The inset shows the typical long-time behaviour of the decay in the quasi-1D traps (red circles) together with the slow decay of immobile molecules in a deep 3D lattice ($\omega_{z,r} = 2\pi \times 17.2$ kHz) (blue squares). $N_0$ denotes the number of molecules at $t = 0$ and the continuous lines are guides to the eye. **b**, Measured longitudinal width $\sigma_z^c$ of the molecular cloud. The continuous lines are damped cosines. All data points are averages over about 20 measurements. The bars indicate the corresponding standard mean error.

Next, we study the inelastic collisions of the Rb$_2$ triplet $v = 0$ molecules which are produced in precisely-defined internal quantum states via coherent optical transfer starting from the Feshbach state (Methods). Figures 4b, c show decay curves of these dimers with rotational quantum number $R = 0$ and $R = 2$, respectively. As for the Feshbach molecules, the observed loss is almost entirely due to collisions since the measured lifetime in the absence of molecular encounters is on the order of several seconds[43]. For direct comparison, we also present in Fig. 4a a data set obtained with Feshbach molecules. Within 5% the laser intensities of the optical lattice are the same for all three data sets. Remarkably, the measurements clearly reveal that the decay of molecules in state $v = R = 0$ takes place on a similar timescale as compared to the $v = 0, R = 2$ molecules or the highly excited Feshbach molecules. This is not obvious because the relaxation paths are potentially different for these states. Specifically, while the Feshbach molecules can vibrationally relax within the triplet potential $a^3\Sigma_u^+$, our $v = R = 0$ molecules are already energetically in the absolute lowest level of the triplet manifold, also with respect to the hyperfine and Zeeman structure (Methods). Thus, in an inelastic or reactive collision of two of our $v = R = 0$ molecules either a Rb trimer must form or at least one of the two dimers must undergo a spin flip towards the singlet electronic ground state. Nevertheless, judging from our measurements presented here, there is no indication for a suppression of the molecular loss rate due to these restrictions, which is an important result of our experiments. These findings go along with theoretical predictions for collisions of *polar* triplet molecules where a spin flip of the electronic state was not suppressed either[44].

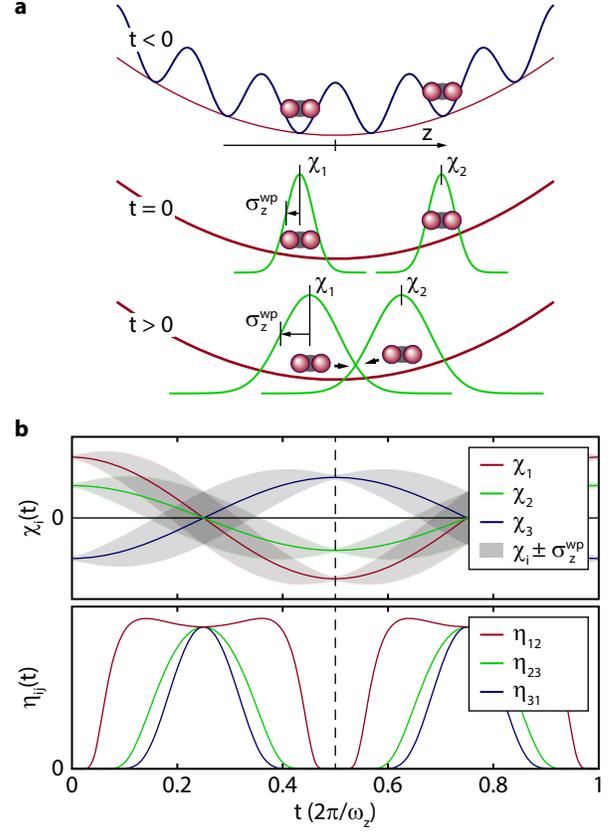

**Figure 3 | Dynamics of molecular wave packets in a quasi-1D potential tube.**
**a**, During the preparation ($t < 0$) the molecules are separated by the deep 3D lattice which suppresses any interaction. At $t = 0$ they are released into the quasi-1D tube as Gaussian wave packets. Their centre positions $\chi_i$ as well as their width $\sigma_z^{wp}$ oscillate with $\omega_z$ and $2\omega_z$, respectively. **b**, For the example of three molecules in a tube we show the dynamics of $\chi_i(t)$, $\sigma_z^{wp}(t)$ (shaded areas) and of the pairwise overlap $\eta_{ij}(t)$.

The step-like loss discussed earlier for the Feshbach molecules is also visible in the data on the deeply bound states, albeit less pronounced. The softening of the steps is caused by smaller initial wave packet widths $\sigma_z^{wp}(t = 0)$ of the $v = 0$ molecules since their polarizabilities at a wavelength of $\lambda = 1064$ nm are by factors between two and three higher as compared to the Feshbach molecules[45]. This leads to a stronger lattice confinement for the same laser intensities and, in addition, to an earlier non-adiabatic release of the wave packets when the lattice is ramped down.

We now investigate the dependence of the reaction rate coefficients on the confinement and collision energy. For this purpose, we measure decay curves for various trap frequencies $\omega_r$ of the tubes. Generally, each of these trap frequencies corresponds to a different collision energy $E_{col}$, because in our current setup we cannot tune $\omega_r$ and $E_{col}$ independently as both are controlled via the laser intensity $I$ of the optical lattice. Specifically, $\omega_r \propto \sqrt{I}$ and the collision energy scales as $E_{col} \propto \omega_r^2 \propto I$ due to the initial potential energies of the particles. The decay rate constants $\overline{\mathcal{K}}_{1D}$ are shown as black circles in Fig. 5 where we use the average trap frequency $\omega_r = \sqrt{(\omega_x^2 + \omega_y^2)/2}$ as a scale for radial confinement (Note that the $v = 0, R = 2$ state exhibits different polarizabilities in $x$- and $y$-direction[45]).

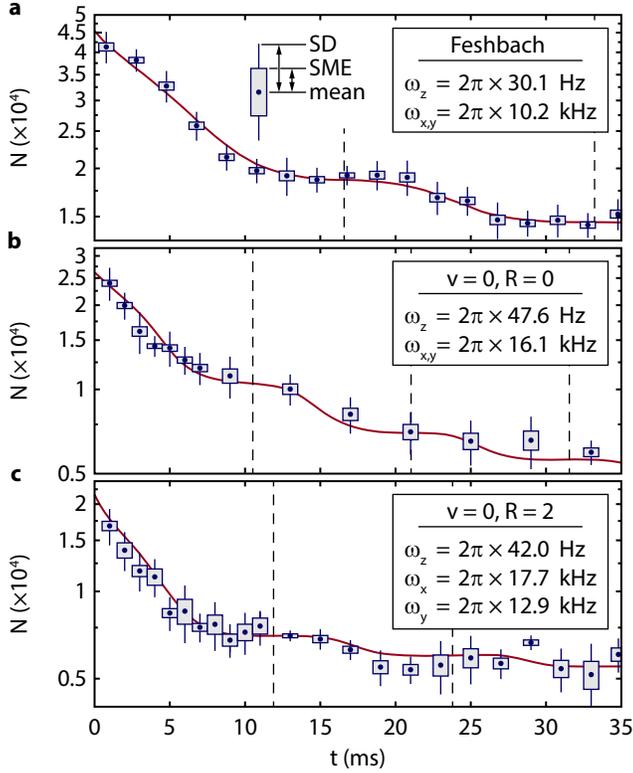

**Figure 4 | Decay curves for various molecular quantum states. a**, Feshbach, **b**, $(v = 0, R = 0)$ and **c**, $(v = 0, R = 2)$ molecules. The laser beam intensities of the 2D lattice are the same within 5% for **a, b, c**, resulting in the trap frequencies provided in the insets. Each data point consists of 10-30 single measurements. In addition to the mean value we give the standard deviation (SD), as well as the standard mean error (SME). The continuous curves are fitted model calculations based on Eq. (1). The dashed vertical lines mark multiples of the time $\pi/\omega_z$, indicating when the cloud size $\sigma_z^c$ is maximal.

We gain additional insights about the dependencies of the rate coefficient from theoretical considerations. For this, we use a quantum defect reaction model where an inelastic process takes place at short range with probability $P_{\text{re}}$[39,40]. Based on the experimental observation that the majority of the molecules is lost in their first encounter with another molecule, quite independently of the initially prepared molecular state, we expect the reaction probability $P_{\text{re}}$ to be close to unity. In the limit that $P_{\text{re}}$ is unity, we are in the universal regime and the resulting s-wave scattering length can be written in a simple form $a_{3D} = \bar{a}(1 - i)$ (see Methods for the general formula with $P_{\text{re}} \neq 1$). Here $\bar{a} = 2\pi R_6/\Gamma^2(1/4)$ is the mean scattering length of the van der Waals potential[46], $\Gamma$ is the gamma function and $R_6 = (2\mu C_6 \hbar^{-2})^{1/4}$ where $\mu$ is the reduced mass of the molecules. The $C_6$ coefficient is $C_6 \approx 17550$ a.u. ($C_6 \approx 18800$ a.u.) for the $v = 0$ molecules (Feshbach molecules), respectively[47]. In free space these parameter values would correspond to universal reaction rate constants[39] $\mathcal{K}_{3D} = 4\hbar\bar{a}/\mu \approx 1.26 \times 10^{-10}$ cm$^3$/s and $1.35 \times 10^{-10}$ cm$^3$/s respectively, which roughly agree with measured reaction rate constants for $^{87}$Rb$_2$ Feshbach molecules[15,24].

The description changes when reducing the dimensionality of the scattering process[35,48]. Generally, a system enters the quasi-1D regime for large trap aspect ratios $\omega_x, \omega_y \gg \omega_z$ and low enough collision energies $E_{\text{col}} < 2\hbar\omega_{x,y}$. For our experiments, we estimate the average maximal energies to be a factor of two to three below this boundary (Supplementary Information, Sec. 4) and $\omega_{x,y}/\omega_z$ is at least $> 300$ for the three investigated molecular states. Thus, the system can be described by an effective 1D model characterized by the interaction potential $V(z) = \frac{\hbar^2}{\mu a_{1D}}\delta(z)$ with complex 1D scattering length $a_{1D}$[49,50] which is a function of $\bar{a}$ and the trap confinement (Methods). Parametrizing $a_{1D}$ as $a_{1D} = (\alpha_{1D} - i\beta_{1D})^{-1}$ allows for writing the universal 1D reaction rate constant in the form[35]

$$\mathcal{K}_{1D} = \frac{4\hbar k^2}{\mu} \frac{\beta_{1D}}{k^2 + \alpha_{1D}^2 + \beta_{1D}^2 + 2k\beta_{1D}}, \quad (2)$$

where $\hbar k$ is the relative momentum of the colliding molecules. The rate constant $\mathcal{K}_{1D}$ can be considered a function of the collision energy $E_{\text{col}}$, since $E_{\text{col}} = \hbar^2 k^2/(2\mu)$. In addition, through the connection between 1D and 3D scattering lengths, $\mathcal{K}_{1D}$ also depends on the transverse confinement $\omega_r$, i.e. $\omega_x$ and $\omega_y$ (see Eq. (6) in Methods). In the low energy limit, $k \to 0$, the one-dimensional rate constant $\mathcal{K}_{1D}$ vanishes as $k^2$, in contrast to 3D where rate constants generally approach a constant value[48]. This is a manifestation of the change of Wigner threshold laws under confinement. In order to compare scaling of the 1D and 3D cases in more detail we examine the decay rates $\Gamma_i = 2\mathcal{K}_i n_i$, where $i \in \{1D, 3D\}$, and $n_{1D}$ represents a 1D density of molecules. In the low energy limit, the ratio $\Gamma_{1D}/\Gamma_{3D}$ is proportional to $(kd)^2 \times d^2/\bar{a}^2 \times (n_{1D}/d^2)/n_{3D}$,[35,48] where $d$ is the transverse confinement size, as given by the harmonic oscillator length (Methods). We note that $n_{1D}/d^2$ can be thought of as an equivalent 3D density of the confined system. The decay rate ratio strongly depends on both confinement and collision energy and indicates that at low enough temperatures and strong confinement the 1D gas would be much more stable than the 3D one.

Figure 5 displays the full dependencies of the universal $\mathcal{K}_{1D}$ on collision energy and confinement as described by Eq. (2) for both the Feshbach and the $v = 0$ molecules. We note, however, that the calculated values of $\mathcal{K}_{1D}$ cannot directly be compared to the experimentally determined values of $\overline{\mathcal{K}}_{1D}$ for two reasons. First, the measured $\overline{\mathcal{K}}_{1D}$ are only given as a function of confinement $\omega_r$ and the corresponding average collision energies are not constant. Second, we have to take into account the dynamics of the molecules and their oscillating energy distribution within the lattice. Therefore, we calculate an approximate theoretical $\overline{\mathcal{K}}_{1D}$ by time-averaging over $\mathcal{K}_{1D}$ according to

$$\overline{\mathcal{K}}_{1D}(\omega_r) = \frac{\sum_{i \neq j} \int \eta_{ij}(t)\mathcal{F}(t)\mathcal{K}_{1D}(E_{\text{col}}(t), \omega_x, \omega_y)dt}{\sum_{i \neq j} \int \eta_{ij}(t)\mathcal{F}(t)dt}. \quad (3)$$

Here, the summations cover all possible colliding pairs of molecules $(i, j)$ in the whole sample with their respective collision energy $E_{\text{col}}$. The theoretical calculations for $\overline{\mathcal{K}}_{1D}$ assuming the universal reaction are shown in Fig. 5 as white solid lines. The universal model reproduces the overall slopes of the data, but generally underestimates the measured decay rate constants. This can be explained by a slight non-universal character of the inelastic collisions for all three investigated states as will be shown in the following.

In order to account for non-universality in our calculations, we use the full expression of $a_{3D}$ as given by Eq. (5) of the Methods section. The theoretical values of $\overline{\mathcal{K}}_{1D}$ are calculated, again using Eqs. (2) and (3). The non-universal model introduces two free fit parameters, the short range parameter $s$ and the reaction probability $P_{\text{re}}$ (Methods). The white dashed lines in Fig. 5 show the results of fits of this non-universal model to the data. For the Feshbach and the $v = 0, R = 0$ molecules good agreement with the measurements is achieved for a wide range of parameter values of $s$ at reaction probabilities $P_{\text{re}}$ between 0.4 and 0.9. The enhancement of the reaction rate relative to the universal regime is due to a shape resonance caused by a near-threshold bound state. For the $v = 0, R = 2$ state the best agreement with the data is obtained when maximizing the calculated rate which is achieved for $P_{re} \approx 0.8$ and a negative $s$. However, still some discrepancy remains. These rotationally excited molecules feature a partial spatial alignment of the molecular axes[45]. Specifically, they collide with their axes pointing dominantly perpendicular to the longitudinal direction of the tubes, which is in contrast to the $R = 0$ molecules where the axis distribution is isotropic. This leads to additional quadrupole-quadrupole interactions. We checked, however, that these interactions are far too weak to significantly influence the reaction rates.

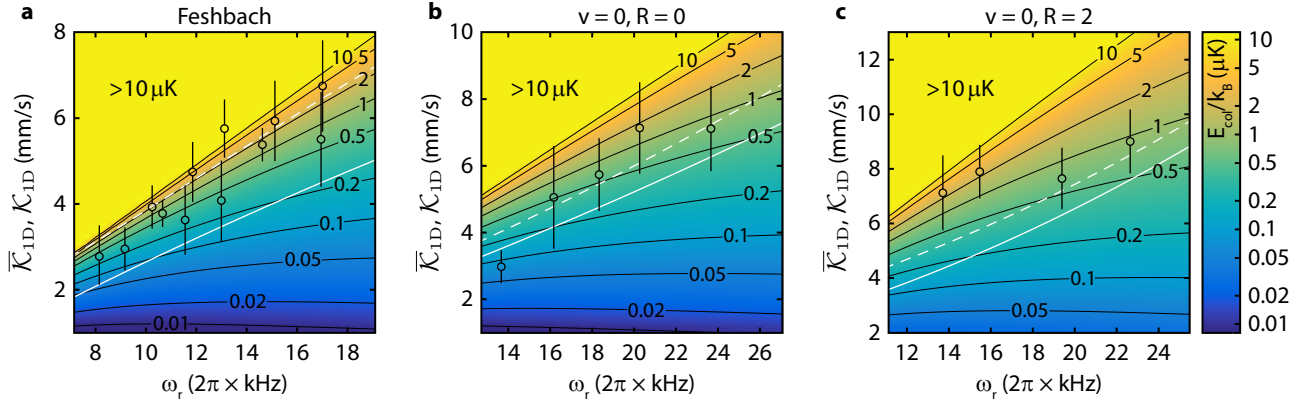

**Figure 5 | Reaction rate coefficients $\mathcal{K}_{1D}$ and $\overline{\mathcal{K}}_{1D}$. a**, Feshbach, **b**, $(v = 0, R = 0)$ and **c**, $(v = 0, R = 2)$ molecules. We show $\mathcal{K}_{1D}$ for the universal case, calculated using Eq. (2). $\mathcal{K}_{1D}$ is a function of both $\omega_r$ and the collision energy $E_{col}$, i.e. $\mathcal{K}_{1D}(\omega_r, E_{col})$. $E_{col}$ is displayed in terms of the background colouring and the equi-energy lines (black solid lines) which are given in units of $\mu K \times k_B$. The black circles represent the values of $\overline{\mathcal{K}}_{1D}$ extracted from the experimental decay curves (as in Fig. 4). We note that $\overline{\mathcal{K}}_{1D}$ is presented here only as a function of the transverse confinement $\omega_r$. As error bars we give the 95% confidence interval of the fits. The solid white lines are the calculations for $\overline{\mathcal{K}}_{1D}$ according to Eq. (3) assuming universality in the reaction model. The white dashed lines correspond to calculations for a non-universal reaction model where the reaction probabilities $P_{re}$ are adjusted to describe best the measured data (i.e. the black circles). Here, the parameter sets for the white dashed lines are $(P_{re} \approx 0.9, s \approx 15)$ for the Feshbach molecules, $(P_{re} \approx 0.9, s \approx 3)$ for the $v = 0, R = 0$ molecules and $(P_{re} \approx 0.8, s \approx -2)$ for the $v = 0, R = 2$ molecules.

In conclusion, we presented the first experimental determination of reaction rates for deeply-bound triplet molecular states. In general, we observe that the majority of the molecules is already lost in the first encounter. We also demonstrate that the decay rate constant in one dimension depends on the confinement strength and the relative energy of the molecules. This offers the possibility to tune the inelastic interaction by adjusting the trap parameters. In this context it could also be insightful to extend the collision experiments to the quasi-2D and 3D regimes since theory predicts that the dimensionality influences the energy dependence of the reaction rates[35–38]. Our measurements indicate that inelastic collisions of $Rb_2$ molecules are characterized by rate constants close to universality. It would be interesting to pursue the investigation further, by mapping out how the rate constant depends individually on the confinement strength and the relative energy of the molecules. Furthermore, stereochemical aspects can be investigated by adjusting the relative alignment of the molecular axes via the preparation of specific rotational states[45]. The presented experimental results together with the provided theoretical model have to be considered as a first step that might pave the way for a fundamental understanding of ultracold chemical reactions and their spatiotemporal control.

## METHODS

**Preparation of cold molecules** The molecules are created in an optical lattice which consists of three perpendicular, retro-reflected laser beams of wavelength $\lambda = 1064$ nm. For this, a thermal sample of roughly $10^6$ $^{87}$Rb atoms at a temperature of about $1$ $\mu K$ is loaded into this lattice such that a signifi-cant number of sites is doubly occupied. The atoms reside in the lowest Bloch band of the 3D optical lattice and the optical potential is deep enough such that tunnelling is strongly suppressed. To first approximation the atoms are normally distributed in configuration space $\propto \prod_{q=x,y,z} \exp\{q^2/(\sqrt{2}\sigma_q^c)^2\}$ with the widths of $\sigma_{x,y,z}^c(t = 0) \approx (26, 27, 26)$ $\mu m$. The atomic occupation of each individual lattice site is described by Poisson statistics. The atoms are in the electronic ground state with the total angular momentum quantum numbers $f = 1$ and $m_f = 1$. By using magnetic Feshbach association at 1007.4 G we create weakly bound $s$-wave molecules at lattice sites occupied with exactly two atoms[41,51]. All remaining atoms are removed in a subsequent purification step[41], such that a pure molecular sample is obtained. The molecular cloud has a size of $\sigma_{x,y,z}^c(t = 0) \approx (22, 24, 23)$ $\mu m$ after production and contains about $4.5 \times 10^4$ Feshbach molecules. The uncertainties for the given values are about 10%. Afterwards, for a large part of our experiments, the molecules are transferred from the Feshbach state to the vibrational ground state $v = 0$ of the energetically lowest triplet potential $a^3\Sigma_u^+$. This is done via stimulated Raman adiabatic passage (STIRAP)[1,45] with a transfer efficiency of about 80%. We prepare the molecules either in one of two quantum levels: The first one is described by the quantum numbers $R = 0, I = 3, F = 2, m_F = 2$ where $I$, $F$, $m_F$ denote the total nuclear spin, the total angular momentum and its projection, respectively. This level is energetically the absolute lowest level of the triplet state, see [52]. The second level is rotationally excited by two units of angular momentum and has the quantum numbers $R = 2, I = 3, F = 4, m_F = 4$. Compared to the $R = 0$ level it has about 2 GHz $\times$ $h$ higher energy, see [52]. The clouds of the $v = 0$ molecules (both $R = 0$ and $R = 2$) are smaller than the cloud of Feshbach molecules, i.e. $\sigma_{x,y,z}^c(t = 0) \approx (19, 20, 18)$ $\mu m$, and their particle numbers range from $2.5 \times 10^4$ to $3.2 \times 10^4$. For the given values the uncertainties are again about 10%.

**Measuring the molecule number and cloud size** In order to measure the total number $N$ and the cloud size $\sigma_{x,y,z}^c$ of the molecules in the 1D tubes at a particular point in time the 3D lattice is quickly switched on. This locks the molecules in their current positions. If they are in a $v = 0$ state we transfer them back to the Feshbach state using STIRAP and subsequently dissociate them by magnetically ramping back over the Feshbach resonance. The resulting atoms are suddenly released from the optical lattice and after a short (200 $\mu s$) time of flight the atom cloud is imaged using standard absorption imaging for another 200 $\mu s$. We note that in our imaging procedure the number of molecules is in general underestimated and the measured cloud size is too large. In the Sup-plementary Information (Sec. 2) we discuss the underestimation of the particle number in detail and derive a correction, which is applied to all our measurements. The overestimated cloud size is a result of the limited resolution of about $4\,\mu m$ of the imaging optics and of the expansion of the atomic cloud during time of flight, and during absorption as well as due to tunneling in the 3D optical lattice shortly before detection. This tunneling is enhanced because a sizeable fraction of molecules will be excited to higher Bloch bands when they are rapidly reloaded back into the 3D lattice. This leads to an overestimation of the molecular cloud size $\sigma^c_{x,y,z}$ of about $3\,\mu m$. Therefore the relative influence is largest on smaller clouds. While we present the originally measured values for $\sigma^c_{x,y,z}$ in the text (see also Fig. 2b), we used corrected values for the simulations.

**Determination of trap frequencies** Both, the radial ($\omega_{x,y}$) and longitudinal ($\omega_z$) trap frequencies are determined from measurements with Feshbach molecules. We use modulation spectroscopy[43] to obtain $\omega_x$ and $\omega_y$, respectively, while $\omega_z$ is inferred from the periodicity of the steps in the molecular decay curves (cf. Fig. 2). The corresponding trap frequencies for the deeply bound $v = 0$ states are derived by comparing their known polarizabilities to the ones of the Feshbach molecules[43,45]. As a consistency check, we find agreement between the so-predicted and the experimentally observed periodicity of the decay steps for both $v = 0$ states (see Fig. 4).

**Numerical integration of the rate equation** In order to numerically integrate Eq. (1) we use the random number generated distribution of molecules (see Fig. 1b), which assigns an initial location $\chi_i(t=0)$ to each particle $i$. Next, we propagate the wave packets of all molecules in small time steps $\Delta t$. The decay probability of each molecule pair $(i, j)$ during $\Delta t$ is given by $\eta_{ij} \overline{\mathcal{K}}_{1D} \mathcal{F} \Delta t$. If an inelastic collision takes place, both involved particles are removed from the sample.

**Scattering in quasi-1D geometry** Since in our setup $R_6 < d_{x,y,z} = \sqrt{\hbar/(\mu\omega_{x,y,z})}$, the collision in the presence of the trap can be described within the pseudopotential approximation with (in general) energy-dependent 3D scattering length[50,53]. In this treatment the interaction potential is replaced by the regularized Dirac delta function and in addition an effective 1D model can be derived[49], using $V(z) = \frac{\hbar^2}{\mu a_{1D}} \delta(z)$. The resulting one-dimensional scattering length $a_{1D}$ is connected to the three-dimensional one via[54]

$$a_{1D} = -\frac{d_y^2(1 - C\,a_{3D}/d_y)}{2a_{3D}\sqrt{\omega_x/\omega_y}}. \tag{4}$$

Here, $C$ is a numerical factor depending only on the transverse trap anisotropy $\omega_x/\omega_y$. For the $R = 0$ state (and the Feshbach state) $C = -\zeta(1/2) \approx 1.46$, where $\zeta$ denotes the Riemann zeta function, while for the $R = 2$ state $C \approx 1.57$. The complex valued 3D scattering length can be written as[40]

$$a_{3D} = \bar{a}\left(s + y\frac{1 + (1-s)^2}{y(1-s) + i}\right). \tag{5}$$

Here $y$ is defined via $P_{re} = 4y/(1+y)^2$, where $P_{re}$ denotes the short range inelastic process probability. The parameter $s$ is the value of the scattering length in the absence of inelastic collisions in units of $\bar{a}$.

The short range reaction probability $P_{re}$ approaches unity in the limit $y \to 1$. Inserting $y = 1$ into Eq. (5), one observes that the scattering length approaches $a_{3D} = \bar{a}(1 - i)$ regardless of the value of $s$. This means that the reaction dynamics becomes universal in the sense that it is independent of the short range details of the potential which normally determine the scattering length. For $y = 1$, Eq. (4) can be rewritten as

$$a_{1D} = -\frac{d_y^2}{2\bar{a}\sqrt{\omega_x/\omega_y}}\left(-\frac{\bar{a}C}{d_y} + \frac{1}{2} + \frac{i}{2}\right). \tag{6}$$

**Acknowledgements**
The authors would like to thank Tommaso Calarco for fruitful discussions and Eberhard Tiemann for valuable information. K.J. is grateful for the hospitality of Tommaso Calarco. This work was supported by the German Research Foundation (DFG) within the project DE 510/2-1, by the Alexander von Humboldt Foundation and by the Foundation for Polish Science within the START programme.

# Supplementary Information
to: Inelastic collisions of ultracold triplet Rb$_2$ molecules in the rovibrational ground state

Björn Drews[1], Markus Deiß[1], Krzysztof Jachymski[2], Zbigniew Idziaszek[2] and Johannes Hecker Denschlag[1]

[1] *Institut für Quantenmaterie and Center for Integrated Quantum Science and Technology IQ$^{ST}$, Universität Ulm, D-89069 Ulm, Germany*
[2] *Faculty of Physics, University of Warsaw, Pasteura 5, 02-093 Warsaw, Poland*

## Contents

## 1 Derivation of rate equation

In the following we derive the rate equation (1) of the main part of the paper. Let $\Psi_i(z,t)$ denote the (normalized) wave function of particle $i$ (in our case $\Psi_i(z,t)$ is a Gaussian wave packet). Then, the wave function of two bosonic particles $i, j$ is

$$\Psi_{ij}(z, z', t) = \frac{1}{\sqrt{\mathcal{N}(t)}} \left( \Psi_i(z,t)\Psi_j(z',t) + \Psi_i(z',t)\Psi_j(z,t) \right), \quad (S1)$$

where $\mathcal{N}(t) = \int dz\, dz'\, |\Psi_i(z,t)\Psi_j(z',t) + \Psi_i(z',t)\Psi_j(z,t)|^2$ is the normalization factor which ensures that the two-particle wave function is normalized for each $t$, i.e. $\int dz\, dz'\, |\Psi_{ij}(z,z',t)|^2 = 1$. Because we are only considering contact interactions here, the probability $d^2 P_R(i,j)$ for a reaction to take place within the infinitesimal intervals $dz$ and $dt$ has to be proportional to the probability of finding both particles within $dz$:

$$\begin{aligned} d^2 P_R(i,j) &= \mathcal{K}_{1D} |\Psi_{ij}(z, z'=z, t)|^2\, dz\, dt \\ &= \mathcal{K}_{1D} |\Psi_i(z,t)|^2 |\Psi_j(z,t)|^2 \mathcal{F}(t)\, dz\, dt, \end{aligned} \quad (S2)$$

where $\mathcal{F}(t) = \frac{4}{\mathcal{N}(t)}$ and $\mathcal{K}_{1D}$ is a rate constant which in general depends on the details of the collision, such as the collision energy. Strictly speaking, the collision energy is not precisely defined here, because of the momentum uncertainty within the wave packets. Thus, $\mathcal{K}_{1D}$ is effectively a mean value of the rate constant.

If more than two particles are present, the total probability for a two-body reaction $d^2 P_R$ is obtained by summing over all particle pairs,

$$\begin{aligned} 2 d^2 P_R &= \sum_{i \neq j} d^2 P_R(i,j) \\ &= dz\, dt \sum_{i \neq j} \mathcal{K}_{1D} |\Psi_i(z,t)|^2 |\Psi_j(z,t)|^2 \mathcal{F}(t). \end{aligned} \quad (S3)$$

Here, the factor of two comes about due to double counting of the pairs. Next, we make an approximation by simplifying Eq. (S3) to read

$$2 d^2 P_R = \overline{\mathcal{K}}_{1D}\, dz\, dt \sum_{i \neq j} |\Psi_i(z,t)|^2 |\Psi_j(z,t)|^2 \mathcal{F}(t), \quad (S4)$$

where $\mathcal{K}_{1D}$ is replaced by an average rate constant $\overline{\mathcal{K}}_{1D}$ which depends on the mean collision energy of the whole ensemble. For our analysis this approximation is quite convenient but not fully justified since in our experiment the collision energy distribution varies as a function of time and space. Still we decided to work with an average rate constant $\overline{\mathcal{K}}_{1D}$, because this reduces the number of fit parameters to a minimum and thus assures that the fits to the data are meaningful. $2d^2 P_R$ corresponds to the expectation value for the particle loss within $dz$ and $dt$,

$$-\langle d^2 N \rangle = d^2 P_R \times 2 + [1 - d^2 P_R] \times 0 = 2\, d^2 P_R, \quad (S5)$$

where we take into account that two particles are lost in each reaction. Integrating over space yields

$$\langle dN \rangle = -\overline{\mathcal{K}}_{1D}\, dt \sum_{i \neq j} \int |\Psi_i(z,t)|^2 |\Psi_j(z,t)|^2 dz\, \mathcal{F}(t), \quad (S6)$$

which is identical to Eq. (1) of the main part of the paper.

Next we show that Eq. (S6) is equivalent to the conventional two-body decay equation for a Bose-Einstein condensate. For this, we make use of Eqs. (S4) and (S5) and we assume that all particles have the same density distribution $|\Psi_i(z,t)|^2 = n_i = n_1$ and thus $\mathcal{F}(t) = 1$. Furthermore, the total particle number $N = \sum_i 1$ is large. Then, one obtains

$$\begin{aligned} \langle d^2 N \rangle &= -\overline{\mathcal{K}}_{1D}\, dz\, dt \sum_{i \neq j} n_1^2 \\ &= -\overline{\mathcal{K}}_{1D}\, dz\, dt\, N(N-1) n_1^2 \\ &\approx -\overline{\mathcal{K}}_{1D}\, dz\, dt\, N^2 n_1^2 \\ &= -\overline{\mathcal{K}}_{1D}\, dz\, dt\, n^2, \end{aligned} \quad (S7)$$

where $n = N n_1$ is the density distribution of the gas. Since $\langle d^2 N/(dz\, dt) \rangle = dn/dt$, it follows

$$\dot{n} = -\overline{\mathcal{K}}_{1D} n^2, \quad (S8)$$

which is the well known rate equation. In addition, we have numerically confirmed that Eqs. (S6) and (S8) yield the same evolution for wave packets propagating in a box potential.

We point out that in a many-body system the decay process may be affected by the correlations present in the gas and its kinetic properties. For example, in the work of Stoof *et al.*[1] reactive losses from a BEC and a thermal gas are discussed. Since the $g^{(2)}$ correlation function of a thermal gas is by a factor of two larger as compared to the one of a BEC (in the limit of vanishing interparticle distances), the reaction rate coefficient is also by a factor of two higher. Specifically, the inelastic loss for a thermal ensemble of molecules can be characterized by $\dot{n} = -\overline{\mathcal{K}}_{1D}^{\text{th}} n^2$, similar to Eq. (S8), however, the corresponding rate coefficient is given by $\overline{\mathcal{K}}_{1D}^{\text{th}} = 2 \times \overline{\mathcal{K}}_{1D}$. Please note that in the present paper all results are given in terms of $\overline{\mathcal{K}}_{1D}$. In this context we would like to mention the work of Dürr *et al.*[2] for the description of a dissipative Tonks-Girardeau gas, where the correlations strongly suppress the decay.

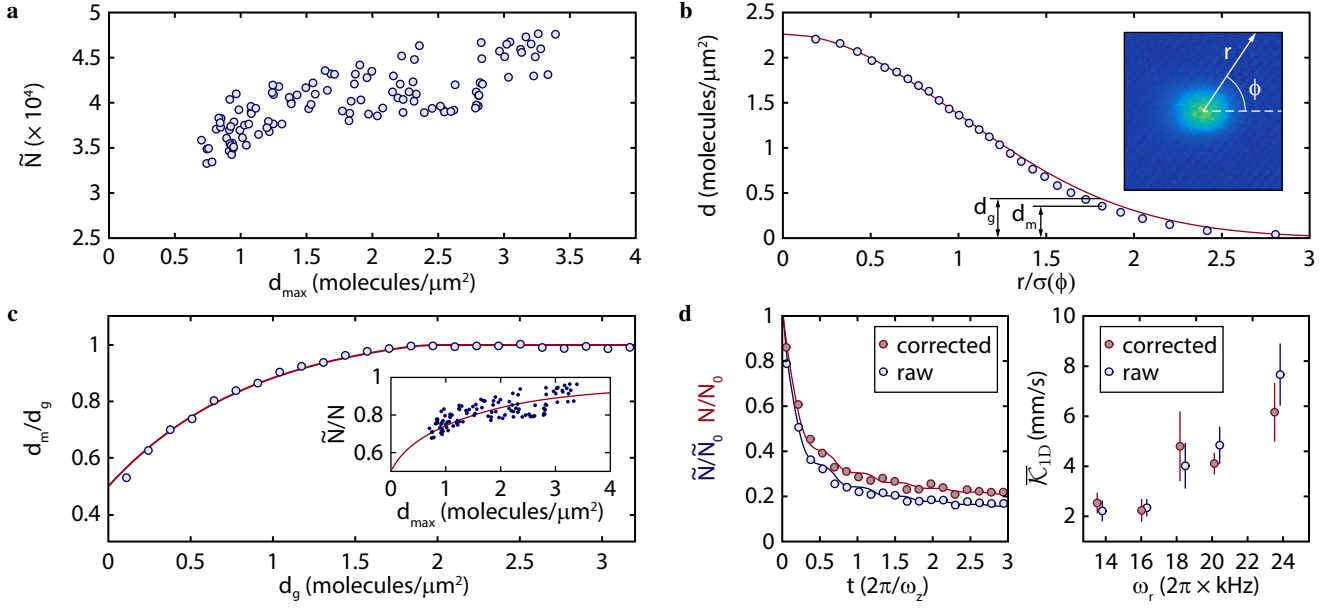

**Figure S1 | Signal loss in the imaging process. a**, Apparent loss in the observed total particle number $\widetilde{N}$ as a function of the density $d_{\max}$ at the centre of the expanding atom cloud. **b**, Deviation of the measured molecular cloud from a Gaussian population distribution. The locally determined density of molecules $d_m$ (circles) is shown versus the (rescaled) radial distance $r/\sigma(\phi)$. It is compared to a Gaussian distribution $d_g$ (solid line), which is fitted to the dense central region of the cloud. The plot represents the azimuthal average (see text) of the image in the inset. **c**, Ratio of the local densities $d_m/d_g$ dependent on $d_g$ with the solid line being the interpolation $f(d_g)$ of the data points. In the inset, we plot the measurements of (**a**) (circles) in terms of the relative particle number $\widetilde{N}/N$, where $N$ denotes the real number of particles expected from our reconstructed density distribution (solid line). **d**, (left) shows a measured decay curve of an ensemble of $v = 0$, $R = 0$ molecules with trap frequencies of $\omega_z = 2\pi \times 40.2$ Hz and $\omega_r = 2\pi \times 13.7$ kHz, either with (red) or without (blue) correction for the signal loss, where the solid lines are the corresponding simulations. The molecule numbers are normalized to their respective initial value at $t = 0$. The plot on the right-hand side summarizes the resulting values of $\overline{\mathcal{K}}_{1D}$ for all data on $v = 0$, $R = 0$ molecules, i.e. different radial trap frequencies $\omega_r$, both for corrected and raw particle numbers. For better visibility the circles are slightly shifted with respect to each other in the horizontal direction.

## 2 Imperfect particle number measurement

As mentioned in the main text, our detection method underestimates atom and molecule numbers at low densities. In the following analysis, we derive a density dependent correction factor which is applied to all measurements in order to obtain the real number of molecules. We first discuss an exemplary measurement: an atomic sample in a harmonic trap is abruptly released for a variable time of flight before it is detected via absorption imaging. During time of flight the cloud expands and its density decreases, while the total number of particles $N$ is conserved. However, as can be seen from Fig. S1a the total number $\widetilde{N}$ obtained from absorption imaging in our setup decreases as a function of the peak (= central) 2D density $d_{\max}$ during expansion. We note that in order to better compare Fig. S1a to the other plots the atom number and density is given in 'molecular units', i.e. $2 \times \widetilde{N}$ is the atom number and $2 \times d_{\max}$ is the atomic density.

The data of Fig. S1a can be used directly to correct the shortcomings of the detection process down to peak molecular densities of about 1 molecule/$\mu m^2$. In order to obtain a correction factor for even lower densities, we use a refined method where we compare the measured density distribution of a cloud of molecules to its known shape. Generally, when a thermal cloud is released from a harmonic trap the expanding density distribution is Gaussian in all three spatial dimensions. Imaging the cloud integrates along one direction and yields a 2D Gaussian distribution. Our measured distributions deviate from a 2D Gaussian. To show this, we first fit a Gaussian function

$$d_g(y, z) = d_{\max} \exp\left( -\frac{(y - \mu_y)^2}{2(\sigma_y^c)^2} - \frac{(z - \mu_z)^2}{2(\sigma_z^c)^2} \right) \quad (S9)$$

to an absorption image (see inset of Fig. S1b). Here, $(\mu_y, \mu_z)$ is the centre position of the distribution and $(\sigma_y^c, \sigma_z^c)$ are the widths of the function along the respective axis. For fitting, only the dense part of the particle distribution is considered (with more than 1.2 molecules/$\mu m^2$). In order to reduce the noise level on our data we carry out azimuthal averaging. For this, we first convert the measured 2D distribution to a circular symmetric 2D distribution by mapping each image pixel with its polar coordinates $(r, \phi)$ to the rescaled coordinates $(r' = r/\sigma(\phi), \phi)$ where the width $\sigma(\phi)$ is determined by $1/\sigma^2(\phi) = \cos^2(\phi)/(\sigma_y^c)^2 + \sin^2(\phi)/(\sigma_z^c)^2$. Afterwards, we bin the data in the $r'$-direction and average over $\phi$ to obtain the radial density distribution $d_m$ as shown in Fig. S1b. If the values of $d_m$ are compared to the radial density distribution $d_g$ of the Gaussian fit, systematic deviations for densities below 1 molecule/$\mu m^2$ are revealed. For each $r' = r/\sigma(\phi)$, we now calculate the ratio between $d_m$ and $d_g$ and average it over more than one hundred images which are obtained for the same initial experimental parameters. These mean values are shown as circles in Fig. S1c while the red curve is an interpolation $f(d_g)$ which reaches essentially unity for densities above 1.8 molecules/$\mu m^2$. In order not to overestimate the signal loss effect, we conservatively set $f(0) = 0.5$. $f(d_g)$ is a transfer function which can be used to calculate the signal loss for the total number of molecules (as observed in Fig. S1a) and to generate a density dependent correction factor. The Gaussian distribution $d_g(y, z)$ of Eq. (S9) is multiplied with the transfer function $f(d_g(y, z))$. Integration yields $\widetilde{N} = \int f(d_g(y, z)) d_g(y, z) dy\, dz$, while $N = \int d_g(y, z) dy\, dz$. Based on this we can determine the ratio $\widetilde{N}/N$ as a function of the observed peak density $d_{\max}$ (see continuous line in the inset of Fig. S1c). As a consistency check we also show the data from Fig. S1a, with $N \approx 5 \times 10^4$ molecules and find good agreement.

Now we are able to correct for the underestimated particle numbers in the measurements. For a given central cloud density $d_{max}$, the measured numbers of molecules are multiplied by the inverse of $\widetilde{N}/N$. Since the Feshbach samples in our experiments exhibit $d_{max} > 2.5 \text{ molecules}/\mu m^2$, the correction factor is comparably low here. In contrast to that, typical central densities for the $v = 0$ molecules are between 1 and $0.1 \text{ molecules}/\mu m^2$ which leads to a relative correction factor of about 1.4 between high and low particle numbers in a set of data. One example is given in the left part of Fig. S1d. It shows the raw data of a molecular decay curve for $v = 0, R = 0$ together with the corrected particle numbers. The step-like structure is partially distorted by the correction process and the heights of the steps become slightly smaller. In Fig. S1d (right) we compare how the decay rate constant $\overline{\mathcal{K}}_{1D}$ changes after correcting the particle numbers. This is done for all our measurements on $v = 0, R = 0$ molecules. Clearly, the change is not very significant and typically lies within the error margins.

## 3 Effects of the STIRAP transfer

The STIRAP transfer of the Feshbach molecules to the $v = 0$ state is not perfect. The overall transfer efficiency is typically about 80%. This produces a fraction of molecules in undetected quantum states (i.e. "dark" molecules) which, however, still contribute to reactive collisions. In addition, the STIRAP transfer is spatially inhomogeneous. This is due to the different dynamical polarizabilities of the Feshbach state and the $v = 0$ states, which in the harmonic dipole trap leads to a spatially varying two-photon resonance frequency. Thus, if the two-photon transition is driven resonantly at the centre of the particle cloud, the energy detuning $\Delta E(x, y, z)$ varies as

$$\Delta E(x, y, z) = \frac{m}{2}(\omega_{v=0}^2 - \omega_{FB}^2)(x^2 + y^2 + z^2), \quad (S10)$$

where $m$ is the molecular mass. $\omega_{FB} = 2\pi \times 44\,\text{Hz}$ is the trap frequency of the Feshbach molecules, while $\omega_{v=0} = 2\pi \times 70\,\text{Hz}$ is the trap frequency for the $v = 0, R = 0$ state. The blue (continuous) curve in Fig. S2a shows $\Delta E(x, y, z)$ along the $z$-axis, i.e. $\Delta E(x = 0, y = 0, z)$. In parallel measurements we have investigated how the STIRAP efficiency depends on the relative detuning of the Raman beams. From these investigations we find that the transfer efficiency can be well described by

$$P(x, y, z) = 0.85 \cdot \exp\left(\frac{-(\Delta E/\hbar)^2}{2\xi^2}\right), \quad (S11)$$

where $\xi \approx 2\pi \times 100\,\text{kHz}$. The factor of 0.85, which represents the maximum efficiency obtained at the centre of the trap, is adjusted to give an overall cloud transfer probability of 80%. Figure S2a shows $P(x = 0, y = 0, z)$ (red dashed line). Its width can be compared to the extension of the initial sample of Feshbach molecules which is illustrated by the shaded distribution.

The imperfect and spatially inhomogeneous STIRAP transfer has a direct influence on the decay signals of the molecules. In Fig.S2b we compare our measurements with the model predictions (which take into account "dark" molecules and their inelastic collisions). For this purpose, we consider the number of remaining observable molecules $N_\infty$ at long times, i.e. after all collisions have taken place. These molecules are trapped in singly occupied tubes. $N_0$ is the total initial number of observable molecules at $t = 0$, i.e. when they are released into the 1D tubes. The data points are directly extracted from the measurements, after subtracting the slow exponential decay due to loss mechanisms other than cold collisions. The dashed (solid) lines are the results of our calculations of the molecular decay for the Feshbach molecules ($v = 0$ molecules). Indeed, the agreement between the measurements and the calculations for the Feshbach state is good, supporting our model for the molecular distribution of Feshbach molecules over the tubes. Now we consider the $v = 0$ states. For a perfectly efficient STIRAP transfer, the data points would fall on the dashed line (which is not the case). If we include the finite transfer efficiency (see Eq. (S11)) of our STIRAP we obtain the solid line which is in agreement with the data points. Similar agreement is found for the $v = 0, R = 2$ state. The reduction in the ratio $N_\infty/N_0$ can partially be explained as follows. As can be seen from Fig. S2a, the inhomogeneous STIRAP transfer is less efficient in the outer parts of the molecular cloud where it dominantly consists of molecules in singly occupied tubes. These account for the main contribution to the $N_\infty$ signal. An imperfect STIRAP transfer reduces $N_\infty$ and thereby $N_\infty/N_0$.

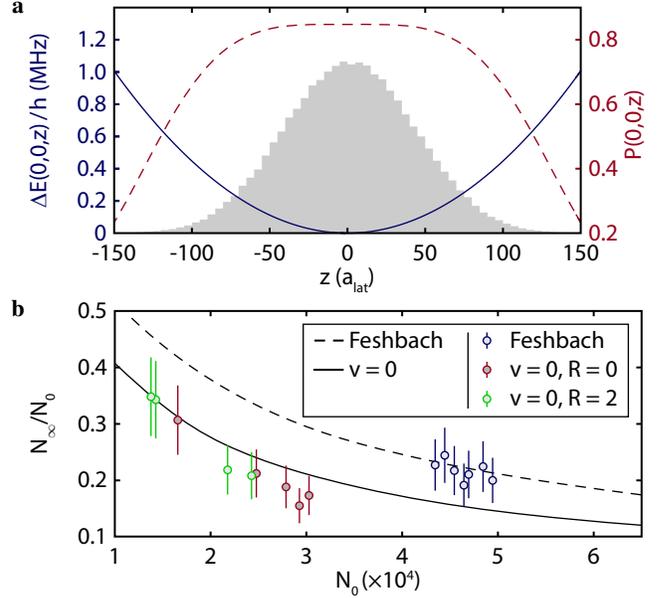

**Figure S2 | Imperfect STIRAP transfer and its consequences. a**, The blue solid line shows the frequency shift $\Delta E(0, 0, z)/h$ of the two-photon STIRAP resonance as a function of the position. Here, $a_{lat}$ is the lattice constant. The frequency shift causes a spatially inhomogeneous transfer probability $P(0, 0, z)$ (red dashed line). For comparison, the shaded area depicts the approximate distribution of Feshbach molecules in arbitrary units. **b**, The ratio of initial ($N_0$) and final ($N_\infty$) molecule numbers is plotted as a function of $N_0$, for both Feshbach and $v = 0$ molecules. The plot symbols show measurements, dashed and continuous lines are corresponding calculations as discussed in the text.

## 4 Damping of cloud oscillations and distribution of collision energies

In Fig. 2b of the main text we observe a damping in the oscillations of the Feshbach molecule cloud size $\sigma_z^c$. Damping should not occur in a true 1D system where only elastic two-body collisions are relevant. Therefore we explain the damping mainly as a consequence of inelastic two-body collisions. As the reaction rate increases with the relative collision energy $E_{col}$, highly energetic molecules are lost faster, which effectively corresponds to damping. Our simulations indeed show, that between 5% and 10% of the amplitude is lost during the first oscillation period. In order to verify the dependence of the damping on the collision energy, we carry out experiments where we measure the damping for various confinements $\omega_z$, since $E_{col}$ scales with $\omega_z^2$. For each $\omega_z$, the damping constant $\gamma$ is extracted by fitting a function $f(t) = a\,e^{-\gamma t}\cos(\omega_z t) + b$ to the data. Figure S3a shows the results for the measurements on Feshbach molecules and reveals a linear dependence of $\gamma/\omega_z$ on the longitudinal trap frequency $\omega_z$, i.e. $\gamma \propto \omega_z^2$. Since $1/\omega_z$ sets the overall time scale for collisions in our setup, the quadratic dependence confirms our explanation.

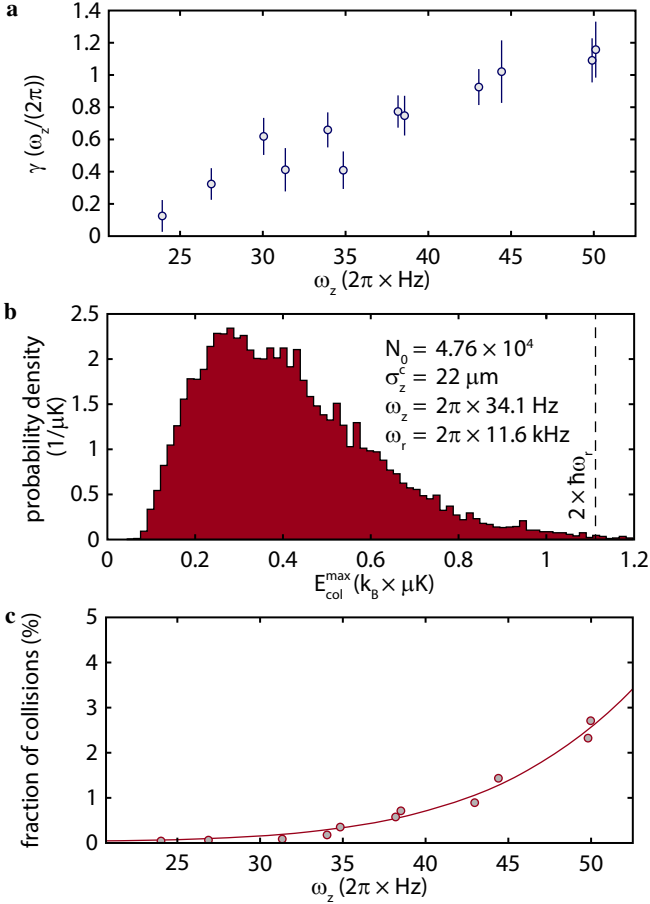

**Figure S3** | **Damping constant and energy distribution. a**, Damping constant $\gamma$ of the oscillations in the width $\sigma_z^c$ of the molecular cloud, extracted from measurements on Feshbach dimers (cf. Fig. 2b of the main text). The error bars represent the 95% confidence intervals of the fits. We note that $\gamma$ is given in units of the inverse oscillation period, so it represents the damping per oscillation. **b**, Normalized histogram of the maximal relative collision energies $E_{\text{col}}^{\text{max}}$ for a sample of Feshbach dimers with the parameters specified in the inset. The dashed vertical line marks the value of $E_{\text{col}}^{\text{max}} = 2\hbar\omega_r$. **c**, Fraction of colliding molecule pairs with $E_{\text{col}}^{\text{max}}$ larger than $2\hbar\omega_r$. The circles are obtained from reconstructed energy distributions using the extracted parameters from the measurements of **a**. The solid line is a guide to the eye.

Finally, we also check that we are well enough in the 1D regime such that elastic two-body collisions do not significantly contribute to the damping. If we were not deeply in the 1D regime, an elastic collision of two molecules could convert part of the collision energy $E_{\text{col}}$ into radial excitation energy $2\hbar\omega_r$ towards the first transverse lattice band. For symmetry reasons such a process requires both particles to be excited to the first band. The transferred portion of energy is missing in the longitudinal motion and thus the oscillation amplitude along this direction is reduced. Assuming a Gaussian distribution of molecules and a known longitudinal trap frequency $\omega_z$, we calculate the distribution of collision energies. For this, we consider for each pair of molecules $i$ and $j$ in a tube its maximal relative energy $E_{\text{col}}^{\text{max}}$. Figure S3b shows a corresponding probability distribution of $E_{\text{col}}^{\text{max}}$ for a typical sample of Feshbach molecules (see figure for cloud parameters).

Noticeably, the fraction of events where the collision energy exceeds $2\hbar\omega_r$ is relatively low. We estimate this by counting the possible molecule collisions with a maximal relative energy of $E_{\text{col}}^{\text{max}} > 2\hbar\omega_r$. According to Fig. S3c the percentage is below 3% for all measurements on Feshbach molecules. Together with the fact that most of the collisions are inelastic, we conclude that non-1D effects in the description of the collisions are negligible.

## 5 Molecular wave packets: size and dynamics

In our model we assume that initially each molecule in the 3D lattice is localised in a single lattice site where it is found in the energetically lowest trap state. It's wave function is then approximately described by a Gaussian wave packet. The dynamics of this Gaussian wave packet after release from the 3D lattice into the 1D tube can be described by the analytical solutions[3] for the wave packet centre $\chi(t)$, the wave packet size $\sigma_z^{\text{wp}}(t)$ and the momentum width $\sigma_p(t)$:

$$\chi(t) = \chi(0)\cos(\omega_z t), \tag{S12}$$

$$\sigma_z^{\text{wp}}(t) = \sigma_z^{\text{wp}}(0)\left[\cos^2(\omega_z t) + \left(\frac{\hbar \sin(\omega_z t)}{2m(\sigma_z^{\text{wp}}(0))^2 \omega_z}\right)^2\right]^{\frac{1}{2}}, \tag{S13}$$

$$\sigma_p(t) = \frac{\hbar}{2\sigma_z^{\text{wp}}(0)}\left[\cos^2(\omega_z t) + \left(\frac{2m\omega_z(\sigma_z^{\text{wp}}(0))^2 \sin(\omega_z t)}{\hbar}\right)^2\right]^{\frac{1}{2}}, \tag{S14}$$

where $\omega_z$ is the longitudinal trap frequency of the 1D tubes. $\sigma_z^{\text{wp}}(0)$ is the initial width and $\chi(0)$ is the initial centre position of the wave packet at $t = 0$, i.e. directly after the release. In the following, we focus on $\sigma_z^{\text{wp}}(t)$ as it is relevant for the description of the molecular decay in our model. Since the initial width $\sigma_z^{\text{wp}}(0)$ cannot be measured directly, we estimate it by numerically calculating how the wave packet in the 3D lattice changes during the ramp-down of the optical potential $U_z$ in the longitudinal direction. This ramp lowers the potential from $U_z^0$ to zero within 400 $\mu$s (see inset of Fig. S4a). The lattice potentials in the $x$- and $y$-directions are kept constant at $U_x^0$ and $U_y^0$, respectively. They determine the trap frequencies $\omega_{x,y,z}$ for the collision experiments in the 1D tubes. In Fig. S4a the red curve displays how the wave packet size $\sigma_z^{\text{wp}}$ at $t = 0$ depends on the initial potential depth $U_z^0$. $U_z^0$ is given in units of the recoil energy $E_r = h^2/(2m\lambda^2)$, where $m$ is the mass of a molecule and $\lambda$ is the wavelength of the lattice laser, i.e. $\lambda = 1064$ nm. A smaller value of $U_z^0$ leads to a larger width $\sigma_z^{\text{wp}}(t = 0)$. This is because a decreasing $U_z^0$ leads to a slower ramp speed for which, in turn, the molecular wave packet can follow the changing lattice adiabatically for a longer time. Thus, the wave packet is released from the lattice with a larger size $\sigma_z^{\text{wp}}(t = 0)$. Our calculations show that for the ramps used in the experiments the wave packet still very much resembles a Gaussian at $t = 0$. This is also revealed when we plot $\hbar/(2\sigma_p(t = 0\mu s))$ (see green line in Fig. S4a), which should be identical to $\sigma_z^{\text{wp}}(t = 0)$ for a Gaussian wave packet. Indeed, for $U_z^0 > 20$ $E_r$ both curves are quite close. For comparison we also show the wave packet size before the ramp-down at $t = -400$ $\mu$s (blue curve).

In our model calculations of the molecular decay we use the calculated values for $\sigma_z^{\text{wp}}(t = 0)$ given in Fig. S4a. The ranges of potential depths $U_z^0$ for our measurements on the Feshbach and the $v = 0$ molecules to extract the reaction rate coefficients are indicated by the grey horizontal bars. For Feshbach molecules this corresponds to a range of $\sigma_z^{\text{wp}}(t = 0) = 0.20 - 0.26$ $a_{\text{lat}}$ while for the $v = 0$ states the values are within $\sigma_z^{\text{wp}}(t = 0) = 0.17 - 0.21$ $a_{\text{lat}}$.

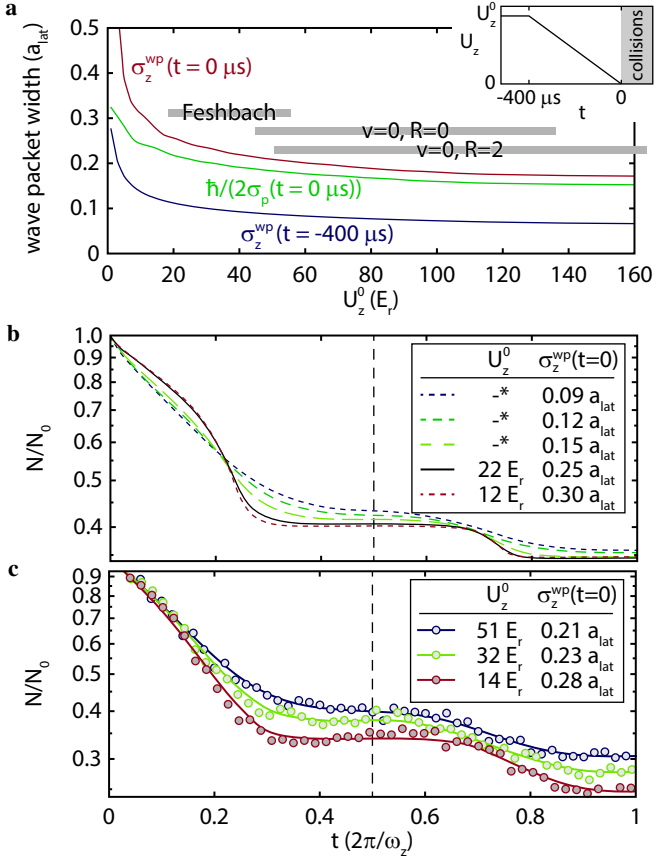

packet widths $\sigma_z^{wp}(t=0) = [0.28, 0.23, 0.21]\, a_{lat}$, respectively.

## 6 Direct calculations of molecular decay curves based on universal model

The rate coefficient given by Eq. (2) of the main part of the publication can be applied to directly calculate the expected molecular decay curves, when we take into account the time-dependence of the oscillating distributions of the molecules in both configuration and momentum space. For this, rate equation (1) has to be modified to read

$$\left\langle \frac{dN}{dt} \right\rangle = -\sum_{i \neq j} \mathcal{K}_{1D}(E_{col}(t), \omega_x, \omega_y) \eta_{ij}(t) \mathcal{F}(t). \quad (S15)$$

As a consequence of assuming universality, there is no free adjustable parameter. Figure S5 shows these full calculations (dashed lines) along with the data and fit curves of Fig. 4 of the main part of the publication. For the calculations we use the same values for $\sigma_z^{wp}$, $\sigma_{x,y,z}^c$ and $N(t=0)$ as for the fit curves. The overall agreement between the data and the universal model calculations is quite reasonable considering the rigidity of the model and the numerous experimental parameters with their respective uncertainties.

**Figure S4 | Wave packet size and influence on molecular decay dynamics.** **a**, Numerical simulation of molecular wave packet widths. The plot shows $\sigma_z^{wp}(t = 0\,\mu s)$ (red) and $\hbar/(2\sigma_p(t = 0\,\mu s))$ (green) of a released molecule after linearly ramping down the optical lattice in the longitudinal direction from the initial depth $U_z^0$ at $t = -400\,\mu s$ to zero at $t = 0\,\mu s$, where the collision experiment starts (see inset). For comparison, we also present the spatial width $\sigma_z^{wp}(t = -400\,\mu s)$ (blue) of the trapped particle in the initial 3D optical lattice. The grey horizontal bars give the range of lattice depths $U_z^0$ for all our collision experiments on the three individual molecular states, which were used to extract the reaction rate coefficients. **b**, Comparison of calculated molecular decay curves for different initial widths $\sigma_z^{wp}(t = 0) = [0.09, \ldots, 0.30]\, a_{lat}$ of the molecular wave packets after release into the 1D tubes. The corresponding values of $U_z^0$ are provided in the inset, where the star symbols mark potential depths that cannot be reached in our setup. For the calculations we used the parameters $N_0 = 4.5 \times 10^4$, $\overline{\mathcal{K}}_{1D} = 120\, a_{lat}/\omega_z$ and $\sigma_{x,y,z}^c(t=0) = (20, 22, 22)\,\mu$m. The dashed vertical line indicates the time of a half oscillation of the molecular cloud. **c**, Measured decay curves of Feshbach molecules. They differ in the ramping speeds (see inset) when releasing the particles into the 1D tubes. The continuous lines are guides to the eye.

In Fig. S4b we present model calculations of molecular decay curves for a range of $\sigma_z^{wp}(t = 0)$ between 0.09 and 0.30 $a_{lat}$. This range is somewhat larger than our experimental one, such that we can display the general dependence of the decay curve on $\sigma_z^{wp}(t = 0)$. Initial small wave packets lead to smoothing out of the decay steps while larger initial wave packets lead to a more pronounced step-like behaviour. However, the relative heights of the steps, which mainly determine $\overline{\mathcal{K}}_{1D}$, do not change considerably. Therefore, a moderate variation in $\sigma_z^{wp}(t = 0)$ has no strong effect on the extracted decay coefficient $\overline{\mathcal{K}}_{1D}$.

We can qualitatively confirm this behaviour in our experiments. Figure S4c shows three measurements for the potential depths $U_z^0 = [14, 32, 51]\, E_r$. According to Fig. S4a these depths correspond to wave

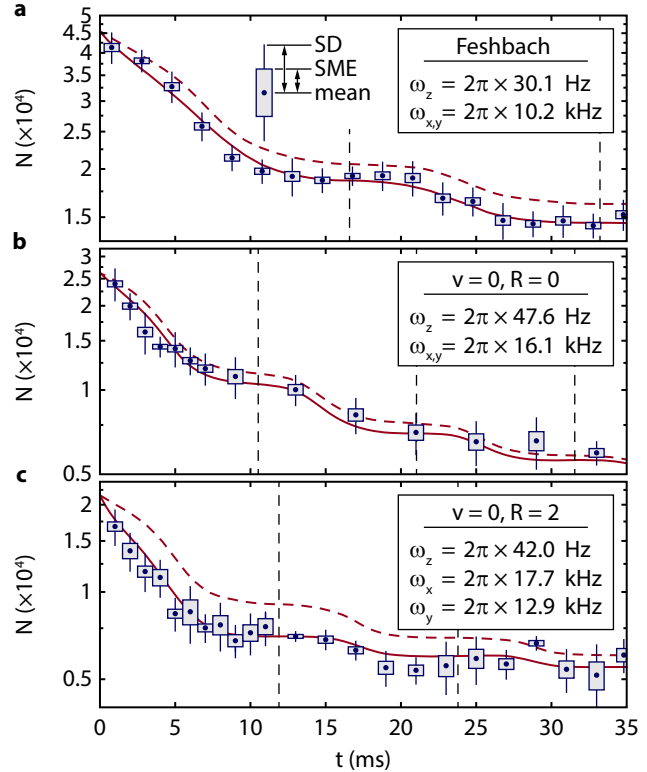

**Figure S5 | Comparison of model predictions to the data for various molecular quantum states.** **a**, Feshbach, **b**, $(v = 0, R = 0)$ and **c**, $(v = 0, R = 2)$ molecules. We show the plot of Fig. 4 of the main part of the publication together with the dashed curves, which are calculations based on Eq. (S15).